# The Structure of Supervoids – I: Void Hierarchy in the Northern Local Supervoid

Ulrich Lindner[1], Jaan Einasto[2], Maret Einasto[2], Wolfram Freudling[3,4], Klaus Fricke[1], and Erik Tago[2]

[1] University Observatory, D-37083 Göttingen, Germany
[2] Tartu Astrophysical Observatory, EE-2444 Tõravere, Estonia
[3] Space Telescope – European Coordinating Facility, D-85748 Garching, Germany
[4] European Southern Observatory, D-85748 Garching, Germany

**Abstract.** Supervoids are regions in the local Universe which do not contain rich clusters of galaxies. In order to investigate the distribution of galaxies in and around supervoids, we have studied the closest example, the *Northern Local Void*. It is defined as the region between the Local, Coma, and the Hercules superclusters, which is well covered by available redshift surveys.

We find that this supervoid is not empty, but it contains small galaxy systems and poor clusters of galaxies. We study the cosmography of this void by analyzing the distribution of poor clusters of galaxies, elliptical and other galaxies in two projections. We present a catalogue of voids, defined by galaxies of different morphological type and luminosity, and analyze mean diameters of voids in different environments.

This analysis shows that sizes of voids and properties of void walls are related. Voids defined by poor clusters of galaxies and by bright elliptical galaxies have a mean diameter of up to 40 $h^{-1}$ Mpc. Faint late-type galaxies divide these voids into smaller voids. The faintest galaxies we can study are outlining voids with mean diameters of about 8 $h^{-1}$ Mpc. Voids located in a high-density environment are smaller than voids in low-density regions.

The dependence of void diameters on the type and luminosity of galaxies, as well as on the large-scale environment shows that voids form a hierarchical system.

**Key words:** cosmology: observations — galaxies: clustering — large-scale structure of the Universe

---

*Send offprint requests to*: Ulrich Lindner, University Observatory, Geismarlandstr. 11, D-37083 Göttingen, Germany

## 1. Introduction

The presence of voids in the distribution of galaxies has been discovered even in early redshift surveys of galaxies (e.g. Jõeveer & Einasto 1978, Tarenghi et al. 1978, Tifft & Gregory 1978 and Tully & Fisher 1978). Further studies show that the largest voids are those delineated by rich clusters and superclusters of galaxies. We call these voids supervoids. This definition is analogous to the definition of superclusters, which consist of rich clusters of galaxies and outline the regions of supervoids. A classical example of a supervoid is the Bootes void (Kirshner et al. 1981). The Bootes void is located at a distance of about 150 $h^{-1}$ Mpc from us. This void is surrounded by the Hercules, Corona Borealis and Bootes superclusters and has a diameter of at least 100 $h^{-1}$ Mpc. Other cluster defined voids are the Perseus void behind the Perseus-Pisces supercluster (Einasto, Jõeveer and Saar 1980) and the supervoid described by Bahcall & Soneira (1982). The closest supervoid – the Northern Local void (hereafter NLV, Einasto et al. 1983) – is located in front of the Hercules supercluster. This void is bordered by the Local and the Coma superclusters and by walls between these superclusters and the Hercules supercluster. From the southern side the NLV reaches the Cetus wall and the Shapley supercluster, cf. Einasto et al. (1994, hereafter EETDA) and Figure 1a.

Catalogues of cluster-defined superclusters and supervoids were presented by Batuski and Burns (1985) and EETDA. The mean diameter of supervoids is about 100 $h^{-1}$ Mpc (EETDA, Einasto et al. 1989, hereafter EEG).

Previous studies have shown that supervoids are not completely empty. In the Bootes void a number of galaxies have been found (Balzano & Weedman 1982, Zmoody et al. 1987, Peimbert & Torres-Peimbert 1992, Szomoru et al. 1993, Szomoru, Gorkom and Gregg 1994, Szomoru et al. 1994). The same has been found for the void near the Perseus-Pisces supercluster (Weinberg et al. 1991), and in the Northern Local Supervoid (Freudling et al. 1988). These studies show that dwarf galaxies in large voids tend to form small systems. In particular, Szomoru et al. have shown, that galaxies in the Bootes void cluster to form groups and filaments of galaxies.

**Fig. 1a.** Distribution of rich clusters of galaxies and strong radio galaxies in supergalactic coordinates. Clusters of galaxies which are members of rich superclusters (with at least 4 members, cf. EETDA) are plotted as filled circles (located in interval $-100 \leq X < 50$ $h^{-1}$ Mpc) or open circles (located in interval $-150 \leq X < -100$ $h^{-1}$ Mpc, different notation enables to distinguish members of the Shapley supercluster), isolated clusters and clusters in poor superclusters are plotted as small dots, radio galaxies as crosses. The zone of avoidance is given by dashed lines. Identifications of some superclusters are given (C–B stands for Corona Borealis, A–C – for Aquarius-Capricornus). Box shows the contour of the region of NLV plotted in Figure 1b.

The aim of the present paper is to investigate the detailed structure of the area of the Northern Local void, which will serve as an example of properties of supervoids to guide future investigations of supervoids. We investigate in detail the properties of voids in the Northern Local supervoid and in the neighboring superclusters – the Local, Coma, and Hercules superclusters. The closeness of this supervoid and the surrounding superclusters gives us a unique possibility to compare structures of galaxies in low- and high-density environment. Differences, if found, give important information on the galaxy formation in various environments.

The Hercules region has attracted the attention of astronomers since Shapley (1934) discovered the Hercules supercluster. The spatial structure of the low-density region in front of the Hercules supercluster was studied by Jõeveer & Einasto (1978), Zeldovich, Einasto and Shandarin (1982), Einasto et al. (1983) and Freudling, Haynes and Giovanelli (1988). This low-density region was independently discovered and studied by Tully (1986, 1987), Tully et al. (1992), by the CfA team (de Lapparent et al. 1986, 1989, Geller and Huchra 1989, Vogeley, Geller and Huchra 1991, Vogeley et al. 1992, 1994a,b), and by Shaver (1991). More recent redshift surveys of the area include Freudling et al. (1992) and Tarenghi et al. (1994). These studies have shown that the region of NLV is filled with a network of faint galaxy systems which divide the supervoid into smaller voids, see Figure 1b. Mean diameters of galaxy-defined voids are 3 – 10 times smaller than diameters of supervoids. These systems are populated mostly by spiral galaxies (Tago et al. 1984, 1986).

**Fig. 1b.** Distribution of galaxies in supergalactic coordinates in a sheet $0 \leq X < 20\ h^{-1}$ Mpc. Bright galaxies ($M \leq -20.3$) are plotted with filled circles, fainter galaxies ($-20.3 < M \leq -19.7$) with open circles. Solid lines show equidensity contours, found by Gaussian smoothing of the density field with dispersion $8\ h^{-1}$ Mpc (cf. Section 4.4). The mean density is denoted as $\bar{\varrho}$. High-density regions ($\varrho \geq 2\,\bar{\varrho}$) correspond to superclusters, low-density regions ($\varrho < \bar{\varrho}$) to the local supervoid, intermediate density regions ($\bar{\varrho} \leq \varrho < 2\,\bar{\varrho}$) to density enhancements in the supervoid, the supercluster outskirts and the walls between them. We see the Local, the Coma and the Hercules superclusters in the lower left, lower right, and upper right part of the figure, respectively. The intermediate density region in the right part of the figure between the Coma and the Hercules superclusters belongs to the Great Wall.

Recently, several catalogues of galaxy-defined voids have been published (Kaufmann and Fairall 1991, hereafter KF, Haque-Copilah and Basu 1992, hereafter HB, and Slezak et al. 1993, hereafter SLB). In these catalogues, however, no distinction has been made between voids defined by different types of objects. By contrast, in this work we emphasize a comparison of properties of voids defined by galaxies of various luminosity and morphology located in different environments. Thereby we confirm previous results by EEG and Einasto et al. (1991, hereafter EEGS), suggesting that the void analysis characterizes some properties of the distribution of galaxies better than other statistics, especially in low-density regions. EEG and EEGS studied the void properties in the direction of the Coma supercluster applying mean void diameter statistics and the void probability function, using redshift data available in the late 80's. In the present paper we shall use a much deeper and larger dataset and combine the quantitative analysis with a cosmographic description of the galaxy distribution.

The paper is organized as follows. In §2 we describe the data used. In §3 we present a cosmographic description of the distribution of galaxies and clusters of galaxies in the direction of the NLV and Coma and Hercules superclusters. In §4 we present a void catalogue in this region of sky. In §5 we analyze the distribution of void diameters defined by poor clusters of galaxies, and by galaxies of different morphological type and luminosity. In §6 we discuss our results. The paper ends with a summary of principal conclusions.

Throughout this paper we use a Hubble constant of $H = h100\ \text{km s}^{-1}\ \text{Mpc}^{-1}$.

## 2. The data

The Hercules supercluster covers a large area of the sky north of the celestial equator in the right ascension range between $12^h < \alpha < 18^h$. Centers of the Local and Coma superclusters lie also in the region between $12^h < \alpha < 18^h$ and $0° < \delta < 30°$. Strong galactic obscuration starts at $\alpha > 18^h$. Therefore we select for our study the region $\delta \geq 0°$ and $12^h \leq \alpha \leq 18^h$. The main body of the Hercules supercluster lies between redshifts 9000 and 12000 km s$^{-1}$. Thus we considered only redshifts up to $\approx$ 12000 km s$^{-1}$. We use cubic samples of clusters and galaxies up to size $L = 120\ h^{-1}$ Mpc, in analyzing data we take into account the fact that in far corners of these cubes data are incomplete.

## 2.1. Galaxies

For galaxies we primarily have used the redshift compilations by Huchra (1991, ZCAT) for the northern sky ($\delta \geq 0°$). Redshift from several sources have been added to improve the completeness in the region of interest. These include several redshift surveys specifically targeting the region of the Hercules supercluster (Freudling et al. 1992, Tarenghi et al. 1994, kindly provided as a computer file by Bianca Garilli). An additional compilation of published redshifts, mostly of spiral galaxies with 21cm line measurements, was kindly provided by Martha Haynes. Finally, unpublished 21cm line measurements in the region of the Hercules supercluster obtained by Marcio Maia and by one of us (WF) were added to the sample. All sources of redshifts were carefully cross correlated to avoid duplicate entries from several sources in our final sample. Only galaxies up to an apparent magnitude of 15.5 have been included in the sample. The resulting sample contains a total of more than 5500 galaxies within the region defined above.

From this combined catalogue we selected a number of subsamples with different magnitude limits and morphological type in order to investigate separately voids defined by all galaxies and by elliptical galaxies (only de Vaucouleurs type $T \leq 0$). A summary of the observational samples used is given in Table 1 (for further explanations of Table 1 cf. Section 5.2).

To assess the completeness of the sample, it was compared to the *Catalogue of Galaxies and of Clusters of Galaxies* (Zwicky et al. 1961-68). It was found that the compilation is complete up to an apparent magnitude 14.5 for almost the whole region. In the declination zones $27° - 50°$ our data are complete up to 15.5. In the declination zones $7.5° - 27°$ and $50° - 75°$ the data are complete up to 15.5 in approximately half of the fields, in the remaining area the completeness varies, and is on average $30 - 50$ %. The data are less complete between $17^h$ and $18^h$. To visualize the completeness of our data we present in Figure 2 histograms of the completeness for each of the Palomar Observatory Sky Survey (POSS) fields located in the direction under study. Because our data are complete up to 14.5 Figure 2 shows the incompleteness in the apparent magnitude range from 14.5 to 15.5.

Since our actual sample is only approximately complete to 15.5 mag, absolute magnitude limited subsamples we use are not strictly volume limited. The impact of the incompleteness on the void analysis is investigated in Section 5.7.

To derive distances from observed redshifts, we have corrected the redshifts for the solar motion (galactocentric flow), the virgocentric flow, and the velocity dispersion in clusters of galaxies (compression) from our list of clusters as described by Einasto et al. (1984), and Einasto et al. (1986).

Distances to galaxies derived from observed redshifts are expressed in redshift space. To compare observed properties of galaxies with models it is necessary either to convert observational data to real space or simulated data to redshift space. Our observational samples are too limited in volume to make the correction to the observational data. In order to estimate this effect on void size statistics we have performed the necessary analysis for model simulations found by Frisch et al. (1994). This analysis shows that the reduction of model samples to redshift space increases mean void diameters only by a few per cent. This is much smaller than statistical errors of void sizes in samples. Thus in the void size analysis we can ignore the difference between redshift and real space.

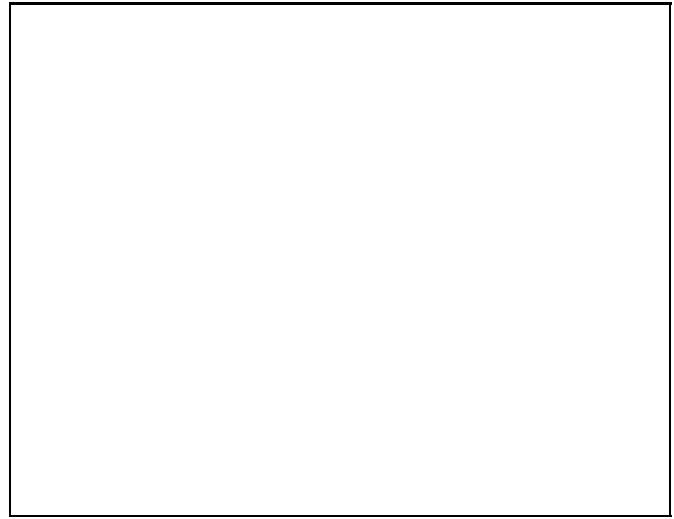

**Fig. 2.** The histograms of the completeness factors in POSS fields, expressed as the fraction of galaxies in our redshift catalogue in units of the number of galaxies in the Zwicky et al. catalogue, for galaxies up to apparent magnitude 15.5. In some fields the fraction is larger than 1, this means that the Zwicky catalogue itself is incomplete.

## 2.2. Clusters of galaxies

In this study we are interested in the distribution of galaxies and systems of galaxies of various richness from poor to rich clusters of galaxies. Thus we used the Zwicky et al. (1961-68) catalogue of clusters which contains data about rich and poor clusters. We used only clusters of the distance class "Near". "Near" clusters are defined as having distances up to 150 $h^{-1}$ Mpc. Most of these clusters contain bright galaxies listed in various compilations of galaxy redshifts (Huchra's ZCAT and others), and therefore it is possible to identify bright members of Zwicky clusters on galaxy catalogues and use them to determine distances of clusters. This study has been carried

**Table 1.** Cubic Samples

| $L$ $h^{-1}$ Mpc | Type | $M_0$ mag | $N(M_0)$ | $\langle D_v \rangle \pm \epsilon_v$ $h^{-1}$ Mpc | $\sigma_v$ $h^{-1}$ Mpc | $N(M \leq -18.8)$ | $N(M \leq -19.7)$ | $N(M \leq -20.3)$ |
|---|---|---|---|---|---|---|---|---|
| 30 | all | −17.3 | 1106 | 8.0 ± 0.6 | 2.4 | 454 | 157 | |
|    | E  |       | 255  | 10.0 ± 1.3 | 3.1 | 128 | 41 | |
| 45 | all | −18.2 | 1345 | 11.0 ± 0.6 | 2.7 | 866 | 282 | 92 |
|    | E  |       | 278  | 15.1 ± 1.7 | 3.5 | 201 | 76 | 28 |
| 60 | all | −18.8 | 1508 | 14.0 ± 0.9 | 4.0 | 1508 | 503 | 146 |
|    | E  |       | 267  | 17.7 ± 1.3 | 4.2 | 267 | 112 | 39 |
| 75 | all | −19.3 | 2114 | 16.0 ± 0.8 | 4.5 | | 1260 | 343 |
|    | E  |       | 323  | 22.0 ± 2.3 | 7.3 | | 223 | 89 |
| 90 | all | −19.7 | 1858 | 18.8 ± 0.9 | 5.3 | | 1858 | 513 |
|    | E  |       | 275  | 27.3 ± 2.3 | 7.8 | | 275 | 115 |
| 105 | all | −20.0 | 1425 | 20.8 ± 0.9 | 5.6 | | | 717 |
|     | E  |       | 215  | 30.3 ± 3.0 | 9.2 | | | 135 |
| 120 | all | −20.3 | 877  | 25.7 ± 1.3 | 6.4 | | | 877 |
|     | E  |       | 141  | 36.4 ± 3.4 | 8.4 | | | 141 |
| 120 | Cluster | | 132 | 37.0 ± 3.5 | 11.7 | | | |

out by Tago (1993) and we used his catalogue of redshifts of Zwicky "Near" clusters of galaxies.

For about 40 per cent of Zwicky "Near" clusters, redshifts were extracted from earlier cluster redshift compilations, or from redshift data for individual galaxies. In other cases redshifts for more than one galaxy per cluster have been obtained from ZCAT. In these cases the unweighted mean of redshifts of individual galaxies in clusters was calculated. In some cases there are several density concentrations in the cluster, in these cases all individual concentrations have were considered as separate clusters located at different distances.

Our sample of the region considered contains 132 Zwicky "Near" clusters of galaxies, 7 of them are also Abell clusters, i.e. they belong to the class of rich clusters.

## 3. Cosmography of Clusters and Galaxies in the NLV region

### 3.1. Visual presentations of the 3-dimensional distribution

Before trying to apply statistical measures to the three dimensional distribution of galaxies, it is useful to get a visual impression first. Therefore, in this section we present two series of plots which show the structure in slices through the region of the NLV.

The first series of plots shows the distribution of galaxies in distance slices in equatorial coordinates (Figure 3). Different symbols are used for different morphological types. Plots are shown for redshift intervals of 1000 km s$^{-1}$.

Another possibility to present the distribution of galaxies and clusters is the use of slices through a rectangular grid. The coordinate system is defined as follows. The positive $x$−axis points towards $\alpha = 12^h$ and $\delta = 0°$, the positive $y$−axis towards $\alpha = 18^h$ and $\delta = 0°$ and the positive $z$−axis towards $\delta = 90°$.

In Figures 4 and 5 cubes with $L = 120\ h^{-1}$ Mpc, and $L = 60\ h^{-1}$ Mpc are shown, respectively. In each Figure, the $z$ coordinate is divided into six equal parts and the resulting sheets are plotted separately. In Figure 4 the distributions of clusters, elliptical galaxies, and all galaxies are given in separate panels. In the cluster panels, Abell clusters are marked with filled circles, and Zwicky clusters with open circles. In Figure 5 we plot only the galaxies, and use different symbols for bins in absolute magnitudes. The highest $z$ panels are not plotted in Figures 4 and 5, they contain much fewer galaxies than other panels and no additional important information.

In Figures 4 and 5 and in the subsequent void size analysis we use absolute magnitude limited data. Over the whole cube the absolute magnitude limit is the same and regions located at different distances from us are represented equally. The absolute magnitude limit corresponds to an apparent magnitude limit of $m = 15.5$ mag at a distance $L$ (the cube sidelength). According to the formula

$$m - M = 15 + 5\log(v) + 0.23/\sin(b) \qquad (1)$$

we get $M = -20.3, -19.7, -18.8$ and $-17.3$ for $L = 120, 90, 60$ and $30\ h^{-1}$ Mpc, respectively. In this formula the redshift $v = cz$ is given in km s$^{-1}$. The term $0.23/\sin(b)$ takes into account the galactic extinction, where $b$ denotes the galactic latitude. In these figures we plot galaxies located in the whole cube, including those in the far corners. Because of this magnitude limit the relevant galaxy data are almost complete in the sphere inside the cube with the center at the origin of the coordinate system and with radius equal to the cube size $L$.

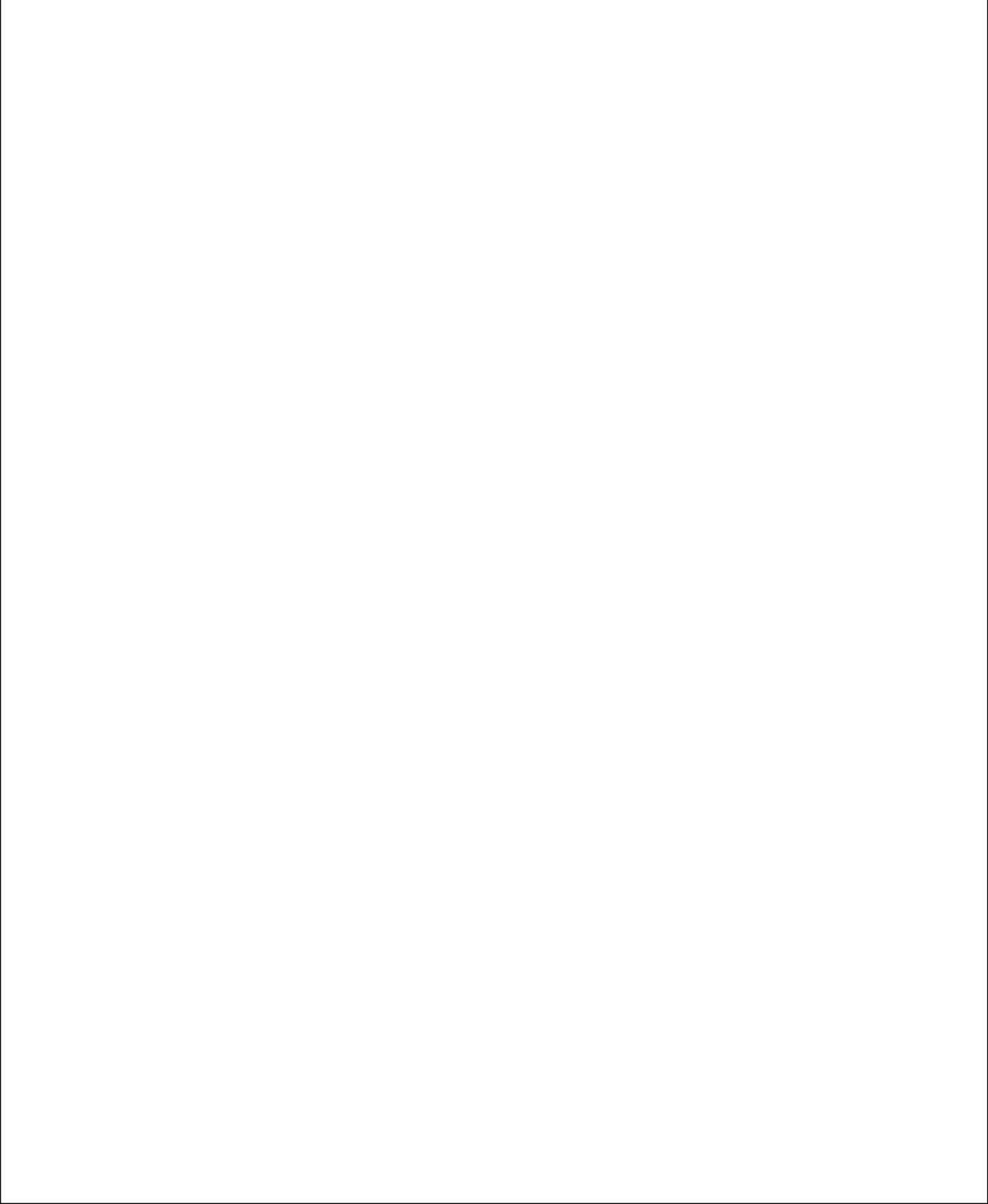

**Fig. 3.** Distribution of galaxies in equatorial coordinates at different distance intervals corresponding to redshift intervals $\Delta cz = 1000$ km s$^{-1}$. Elliptical galaxies are plotted with filled circles, spiral and irregular galaxies with open circles.

**Fig. 4.** Distribution of clusters and galaxies in rectangular equatorial coordinates with cube size 120 $h^{-1}$ Mpc. Cluster distribution is given in left panels, that of elliptical galaxies in middle panels, and of all galaxies in right panels. The $z$ coordinate of the cube is divided into intervals of 20 $h^{-1}$ Mpc. In cluster panels Zwicky clusters are plotted with open circles, Abell clusters with filled circles. In the rightmost panels we show also voids from our void catalogue in Table 2a. The circles are intersections of the spherical voids with planes defined by z=10, 30, 50 and 70 $h^{-1}$ Mpc, respectively.

**Fig. 5.** Distribution of galaxies in rectangular equatorial coordinates with cube size 60 $h^{-1}$ Mpc. The $z$ coordinate of the cube is divided into intervals of 10 $h^{-1}$ Mpc. In left panels only bright galaxies ($M \leq -20.3$) are plotted, in middle panels bright and intermediate bright galaxies ($-20.3 \leq M \leq -19.7$), in right panels galaxies of all three luminosity class are plotted (faint galaxies ($-19.7 \leq M \leq -18.8$). Bright, intermediate and faint galaxies are plotted with filled, large and small open circles, respectively.

Data on cubic samples built up according to the method described above are compiled in Table 1: $L$ is the side length of the cube given in $h^{-1}$ Mpc and $M_0$ the corresponding absolute magnitude limit $M$ according to the formula above. $N$ is the number of galaxies in the respective cubic sample. This number depends on $M_0$; in the following analysis we use samples of galaxies brighter than the limiting absolute magnitudes $-20.3$, $-19.7$ and $-18.8$. The mean void diameter $\langle D_v \rangle$ will be explained in Section 5.1.

### 3.2. Superclusters of galaxies

In the volume under study there are three superclusters – the Virgo, Coma and Hercules superclusters.

In the panel ($1000 < cz \leq 2000$ km s$^{-1}$) of Figure 3 we see in the lower right corner the Virgo cluster, the center of the Local Supercluster. The Virgo cluster is so large on the sky that it has not been included into the Zwicky or Abell catalogue of clusters of galaxies. In the next distance interval there are much fewer galaxies. This marks the outskirts of the Local supercluster. The Local supercluster is also well recognizable in the lower $z$ panels of Figures 4 and 5 near the coordinates $x = 13$ and $y = 3$ $h^{-1}$ Mpc.

In the 6000 km s$^{-1}$ $< cz <$ 7000 km s$^{-1}$ and 7000 km s$^{-1}$ $< cz <$ 8000 km s$^{-1}$ panels of Figure 3 we can recognize the Coma Supercluster. The core of the supercluster is formed by the rich Abell cluster A1656, which is visible as an elongated cluster in the panel $7000 < cz < 8000$. In Figures 4 and 5 the Coma Supercluster is seen at coordinates $x = 60$ $h^{-1}$ Mpc and $y = 15$ $h^{-1}$ Mpc in $20 < z < 40$ and $30 < z < 40$ panels.

Galaxy enhancements cover most of the sky in panels on distance intervals between 8000 km s$^{-1}$ and 12000 km s$^{-1}$ of Figure 3. In these distance intervals we see the Hercules supercluster. It covers a very large area on the sky, and is characterized by a number of rich clusters of galaxies (cf. the list of supercluster members in EETDA). In Figure 4 the Hercules supercluster is seen at $x = 60$ $h^{-1}$ Mpc and $y = 90$ $h^{-1}$ Mpc(panel $20 < z < 40$), here Abell clusters A2152, A2151 and A2147 are located as well as a number of galaxy systems.

In Figure 1a we also show the distribution of bright radio galaxies. These galaxies are located only in high-density environment, indicating the presence of superclusters. A high concentration of radio galaxies to nearby superclusters was noticed by Jõeveer, Einasto and Tago (1978) and Shaver (1991).

### 3.3. Walls between superclusters

Figures 3, 4 and 5 show a number of systems of galaxies which form walls between superclusters.

In the 5000 km s$^{-1}$ $< cz <$ 6000 km s$^{-1}$ panel of Figure 3 a long chain of galaxies can be seen extending from $14^h$ and $30°$ to about $16^h$ and $40°$. This chain forms a connection between the Local and Coma superclusters (Tago, Einasto and Saar 1986).

In the 6000 km s$^{-1}$ $< cz <$ 8000 km s$^{-1}$ panels of Figure 3 chains of galaxies extend from the Coma cluster to left and up (increasing $\alpha$ and $\delta$) – these galaxies belong to the Great Wall (Geller and Huchra 1989). The Great Wall is clearly seen in Figure 1b as a strong chain of galaxies between the Coma and Hercules superclusters.

On distance intervals between 8000 km s$^{-1}$ and 12000 km s$^{-1}$ of Figure 3 (the Hercules supercluster region) clusters of galaxies are connected by systems of giant galaxies of various richness and shape.

In the lower left panel ($0 < z < 20$) of Figure 4 we see a very distinct chain of poor clusters of galaxies extending from the Local Supercluster at the origin of the coordinate system to the Abell Cluster A2063, a member of the Hercules Supercluster.

These figures suggest that walls between superclusters do not form large structure-less clouds of galaxies, but consist of numerous smaller systems of galaxies. Ramella et al. (1992) came to a similar conclusion. These systems are less rich than similar systems in superclusters but contain more galaxies than systems in the supervoid region discussed in the next section.

### 3.4. Systems of galaxies in the NLV

In the region of the NLV, between the Hercules, the Local and the Coma superclusters and the walls between these superclusters, there is a significant number of small systems of galaxies and poor clusters. The distribution of galaxies in the supervoid region is discussed in the following.

In the first panels of Figure 3 (up to redshifts 2000 km s$^{-1}$) we see the main body of the Virgo cluster. Starting from redshifts 1500 km s$^{-1}$ a number of small systems appear, which consist mainly of faint galaxies ($M > -18.8$); brighter galaxies can be seen in the group of galaxies at $\alpha \approx 14^h$, $\delta \approx 40°$ and redshift interval $2000 - 3000$ mentioned, divide the region into a number of voids, in which we do not see any galaxies.

At the lower right corner of the panel $6000 - 7000$ km s$^{-1}$ of Figure 3 we see galaxies forming a wall between two voids (see corresponding region on the neighboring panels). This wall begins at the Coma supercluster area and crosses with another cloud of galaxies at $\delta \approx 10°$. A comparison of different panels shows that this wall consists of a thin net of filaments of faint galaxies. These filaments are thinner than the 1000 km/s slices shown here and therefore can be seen with high contrast when slices of 500 km/s depth are plotted.

In Figure 4 we plot the brightest galaxies ($M \leq -20.3$) and the elliptical galaxies with the same absolute magnitude limit. Giant galaxies show the strongest features in the distribution of galaxies. The distribution of elliptical galaxies is similar to that of giant galaxies: elliptical galaxies are located in denser regions and are absent from low-density regions which consist mainly of fainter later type galaxies.

In Figure 5 the cube size is half the cube size of Figure 4 and consequently the apparent magnitude limit allows to see fainter galaxies. In Figure 5 galaxies of all morphological types with $M \leq -18.8$ are plotted in three panels for each $z-$sheet. In the left panel galaxies are restricted to absolute magnitudes brighter than $M = -20.3$. Those panels correspond to the right panels in Figure 4 apart from the different thickness of the $z-$sheets. In the middle and right panels galaxies for fainter magnitude limits are plotted. Different symbols are used for different absolute magnitude intervals. A comparison of different panels for the same $z-$sheet shows that not all of the faint galaxies are members of structures formed by bright galaxies. Instead, some of the fainter galaxies form new structures which consist only of faint galaxies. A good example is a system of galaxies around the void in the $10 < z < 20$ panel with its center at about $x = 35$ and $y = 40\ h^{-1}$ Mpc. This structure is completely absent in the leftmost panel (galaxies with $M \leq -20.3$) of the Figure. Many other less pronounced but clearly recognizable voids and galaxy systems can be found by comparing corresponding panels in Figure 5. This is a direct indication that structures depend on the luminosity of galaxies: poor systems consists only of faint galaxies.

All these plots show that the NLV region contains a network of galaxies and poor clusters of galaxies. We see both luminosity segregation between giant and normal galaxies and large-scale morphology segregation between elliptical and later type galaxies in systems far from supercluster regions (Einasto 1991, Mo et al. 1992). Faint galaxies tend to lie either in structures already outlined by brighter galaxies, or form new structures. They do not form a smooth population of isolated galaxies in the voids. Our figures do not show the distribution of extremely faint dwarf galaxies, but other studies (Szomoru et al. 1994, Hopp 1994 and references therein) have shown that up to now the search of dwarf galaxies in voids has given negative results.

To summarize this section, the comparison of the distribution of all galaxies, elliptical galaxies, and Zwicky clusters shows that most voids defined by clusters are seen in the distribution of galaxies as well. Clusters form knots in galaxy systems. The richest clusters are located in the region of the Hercules and Coma superclusters, galaxy systems inside the supervoid contain only poor clusters. We see a *hierarchy* of systems of galaxies from poor systems to superclusters in which the morphology and luminosity of galaxies are related to the properties of the systems in which they are located. The void sizes in this hierarchy depend on the objects which have been used to detect the void: voids, defined by clusters are larger than galaxy defined voids, and voids formed by elliptical galaxies are larger than voids defined by galaxies of all types.

These observations motivated the more quantitative approach of Section 5, where we strengthen these impressions by calculations of mean void diameters. The void hierarchy will be quantitatively confirmed by generating void catalogues for galaxy samples of different magnitude limit.

## 4. Distribution of Voids

### 4.1. Method to find voids

The purpose of our void catalogue is to help the study of the dependence of void parameters on the type of objects used in the void definition. The catalogue could also be useful as a basis of searching faint dwarf galaxies in the vicinity of voids. Voids can be defined in a number of ways, and their properties depend on this definition. For this investigation, we define voids as regions which are completely empty of certain types of objects. We used a computer-based void-searching algorithm to determine voids objectively. Topological voids are connected, here we consider geometrical voids.

Three different algorithms have been used in the literature to determine voids, namely the empty cube method by KF, the empty sphere method by EEG, and the smoothed density field method using a wavelet technique by SLB. The first method finds completely empty cubic regions, the third method finds low density regions. Empty sphere and cube methods yield very similar results, only the sizes of voids and void center coordinates are slightly different. The empty sphere method is more sensitive to fine details of the under-dense regions, and numerical results reflect the void structure better. The smoothed density field method is less sensitive to fine details in the distribution of galaxies in low-density regions. In particular, relatively isolated galaxies influence the smoothed density field much less than parameters of completely empty voids, thus the preference depends on the purpose of the study – either the study of under-dense or of empty regions. In our attempt to confirm the visual impression described in Section 3, we are mostly interested in completely empty regions and thus give preference to the empty sphere method.

The principal difference of the present study from previous ones lies in the fact that we take into account the luminosity, morphological type, and environment of galaxies while only little attention has been given to these aspects before.

Visual inspection of Figures 3 – 5 shows that the shapes of voids resemble irregular ellipsoids. For simplicity we identify voids by searching for empty spheres. In the following we use the term *void* synonymous with *empty sphere* keeping in mind that our voids are only a more or less rough approximation to the real (or physical) voids which are very complex and irregular empty regions in space. To find voids in a cubic sample we use the empty sphere method by EEG. In this method we divide our cubic samples into $k^3$ cubic cells, where $k$ is the resolution parameter. The size $l$ of the edge of such a cell is $l = L/k$, where $L$ is the size of the whole cubic sample. First we calculate the distance to the nearest object (galaxy or cluster) for the centers of all cells, and then find local maxima in this distance field. Maxima are located in centers of empty spheres, respective distances correspond to radii of empty spheres. Void diameters determined with this method describe the extent of the smallest axis of a void.

Several problems connected with the application of the empty sphere method have to be taken into account. First, empty spheres found near the borders of the cubic sample may not correspond to centers of real voids because objects located outside the cube are not taken into account. For this reason we define a *reduced* cubic sample where all those empty spheres that lie nearer to any cube border than the radius of the particular void are discarded. Also we discard all voids with centers outside the sphere of radius $L$ around the observer.

We investigated the dependence of our results on the resolution parameter $k$. The number, center position and diameter of voids found depends on this parameter. Typical cell sizes are in the range of about 1 to 7.5 $h^{-1}$ Mpc and typical void radii are in the range of about 7 to 20 $h^{-1}$ Mpc. Voids with radii much greater than the cell size may be counted more than once (especially in the case of voids of irregular shape we may obtain several voids with close centers). The smaller the cell size $l = L/k$, the more often the same physical void may be counted. To reduce multiple void counting we apply an elimination algorithm using the overlapping degree defined in the next section.

### 4.2. The overlapping degree

For further studies we define the overlapping degree $\theta$ of spheres as follows:

$$\theta = \begin{cases} 0, & r_1 + r_2 - a \leq 0; \\ \dfrac{r_1 + r_2 - a}{2min(r_1, r_2)}, & 0 < r_1 + r_2 - a < 2min(r_1, r_2); \\ 1, & r_1 + r_2 - a \geq 2min(r_1, r_2), \end{cases} \quad (2)$$

where $r_1$ and $r_2$ are the radii of the respective spheres and $a$ is the distance between their centers. An overlapping degree $\theta$ equal to zero means that the spheres under consideration are isolated or touching. An overlapping degree $\theta = 1$ means that the smaller sphere is completely inside the larger one. The normalization by the factor of $2min(r_1, r_2)$ (the diameter of the smaller sphere) is used to get values between 0 and 1 as a characteristic number describing the degree of overlapping. We give overlapping in per cent i.e. as numbers from 0 to 100. For example, an overlapping of 50 % means that the center of the smaller sphere lies exactly on the surface of the larger sphere.

### 4.3. The void catalogue

With our void catalogue we intend to cover as much empty space with a reasonable small number of empty spheres as possible. According to our definition of voids this empty space depends on the type of objects considered. For our further studies we use galaxies of all morphological types but of three different absolute magnitude limits to define voids. The void catalogue was constructed in several steps as follows.

In the first step voids were determined with the *empty sphere method* applied to cubic samples of size $L = 60, 90$ and 120 $h^{-1}$ Mpc and corresponding absolute magnitude limit $M = -18.8, -19.7$ and $-20.3$. Resolution parameters $k = 16, 24, 32, 40, 48, 56$ and $64$ have been used. All the voids found for different $L$ and $k$ but with identical $M$ were assembled into three lists, one for each absolute magnitude limit. The purpose of this procedure is to become independent of the resolution parameter. The voids in those compilations were sorted according to their size in descending order and thus we get three *void search lists* for our further proceeding. Note that this list contains a large number of more or less strongly overlapping voids.

In the next step we use this list to find a catalogue of non-overlapping voids which we will call "parent" voids to simplify terminology for further explanations. The first (and largest) void in the void search list is taken as the first "parent" void. Then we go through this list until we meet the next void which does not overlap with the first "parent" void, this is the second "parent" void. Then we go further through the list until we meet the next void which does not overlap with the two preceding "parent" voids and so on. With this procedure we compile a catalogue of all largest non-overlapping voids from the void search list.

All the remaining voids are overlapping with one or more "parents" with different degree $\theta$. We call these voids "children" and each of them will be assigned to that "parent" void where the overlapping degree is largest. Thus "parent" voids together with their "children" are forming "families" of voids. "Parent" voids do not cover the whole empty space, part of it is covered only by "children" voids. To increase the volume of empty space covered by our void catalogue we include some "children" voids to the catalogue by the following procedure. The largest void among the "children" of each "family" was identified, and in case when its diameter was considerably larger (by a factor of 1.3 or more) than the diameter of the "parent" void and if the overlapping with the "parent" void was not very strong (less than 30 %) this largest "child" has been considered as a new "parent" and was added to the void catalogue. Note that these additional voids may overlap with several "parent" voids, but with a smaller $\theta$. All the other "children" of the "family" were taken into account through the ellipticity of the "parent" void (see below).

**Table 2a.** Void Catalogue ($M \leq -20.3$)

| No. | $\alpha$ | | $\delta$ | $R$ | $D_v$ | $e$ | $\varrho$ |
|---|---|---|---|---|---|---|---|
| | h | m | | $h^{-1}$ Mpc | $h^{-1}$ Mpc | | |
| A1  | 12 | 36.4 | 47.0 | 82.4  | 17.3 | 1.55 | 1.82 |
| A2  | 12 | 44.0 | 58.9 | 92.8  | 17.4 | 1.95 | 1.51 |
| A3  | 12 | 49.6 | 32.8 | 80.9  | 14.7 | 1.84 | 4.04 |
| A4  | 12 | 58.8 | 19.6 | 97.8  | 35.1 | 1.47 | 0.21 |
| A5  | 13 | 16.4 | 39.1 | 96.9  | 26.1 | 1.92 | 0.91 |
| A6  | 13 | 20.8 | 24.0 | 33.2  | 20.6 | 1.80 | 1.75 |
| A7  | 13 | 30.8 | 28.6 | 107.7 | 10.7 | 1.00 | 1.69 |
| A8  | 13 | 48.0 | 7.3  | 69.3  | 16.1 | 1.03 | 2.46 |
| A9  | 13 | 51.6 | 59.3 | 90.7  | 21.3 | 1.71 | 1.31 |
| A10 | 13 | 52.4 | 48.9 | 118.1 | 30.4 | 1.59 | 0.24 |
| A11 | 14 | 0.0  | 35.1 | 96.4  | 20.1 | 1.84 | 1.17 |
| A12 | 14 | 8.8  | 28.1 | 67.8  | 27.2 | 1.64 | 1.12 |
| A13 | 14 | 15.2 | 18.7 | 108.4 | 30.0 | 1.79 | 0.91 |
| A14 | 14 | 21.2 | 32.5 | 90.0  | 14.4 | 2.10 | 1.31 |
| A15 | 14 | 30.4 | 7.8  | 118.7 | 28.0 | 1.50 | 0.73 |
| A16 | 14 | 45.6 | 12.4 | 81.6  | 13.8 | 1.35 | 1.93 |
| A17 | 14 | 54.8 | 7.5  | 104.3 | 14.8 | 2.70 | 2.77 |
| A18 | 14 | 55.2 | 35.3 | 118.2 | 37.1 | 1.63 | 0.20 |
| A19 | 15 | 0.0  | 50.4 | 77.4  | 35.6 | 1.57 | 0.36 |
| A20 | 15 | 7.2  | 66.3 | 105.5 | 28.8 | 1.51 | 0.69 |
| A21 | 15 | 21.2 | 10.4 | 117.3 | 21.5 | 1.65 | 2.51 |
| A22 | 15 | 21.6 | 19.8 | 101.9 | 22.5 | 1.63 | 2.02 |
| A23 | 15 | 46.0 | 51.5 | 106.6 | 24.8 | 1.68 | 1.08 |
| A24 | 15 | 47.2 | 9.5  | 97.3  | 18.4 | 1.48 | 3.64 |
| A25 | 16 | 2.0  | 18.0 | 70.5  | 39.5 | 1.61 | 0.08 |
| A26 | 16 | 8.0  | 17.0 | 118.8 | 12.8 | 1.51 | 6.59 |
| A27 | 16 | 8.4  | 35.7 | 95.1  | 7.8  | 1.87 | 7.06 |
| A28 | 16 | 8.8  | 22.2 | 106.5 | 16.5 | 2.31 | 6.55 |
| A29 | 16 | 10.8 | 45.2 | 102.1 | 17.0 | 2.12 | 4.50 |
| A30 | 16 | 22.4 | 66.4 | 72.8  | 14.6 | 1.40 | 1.69 |
| A31 | 16 | 34.0 | 29.9 | 103.4 | 18.2 | 2.10 | 3.81 |
| A32 | 16 | 36.4 | 21.3 | 94.8  | 27.5 | 1.60 | 1.65 |
| A33 | 17 | 8.0  | 52.5 | 117.5 | 31.1 | 1.42 | 0.26 |
| A34 | 17 | 10.0 | 19.8 | 110.8 | 31.9 | 1.70 | 0.53 |
| A35 | 17 | 11.2 | 7.5  | 93.8  | 18.8 | 1.00 | 0.72 |
| A36 | 17 | 15.6 | 35.4 | 116.5 | 27.5 | 1.62 | 0.90 |
| A37 | 17 | 16.0 | 51.6 | 88.8  | 21.1 | 1.79 | 1.36 |

**Table 2b.** Void Catalogue ($M \leq -19.7$)

| No. | $\alpha$ | | $\delta$ | $R$ | $D_v$ | $e$ | $\varrho$ |
|---|---|---|---|---|---|---|---|
| | h | m | | $h^{-1}$ Mpc | $h^{-1}$ Mpc | | |
| B1  | 12 | 12.4 | 22.7 | 71.0 | 6.8  | 1.00 | 3.66 |
| B2  | 12 | 35.2 | 13.6 | 22.0 | 3.4  | 1.00 | 3.23 |
| B3  | 12 | 38.4 | 28.5 | 72.2 | 5.4  | 1.37 | 4.30 |
| B4  | 12 | 38.8 | 28.9 | 89.1 | 11.7 | 1.25 | 1.21 |
| B5  | 12 | 40.8 | 23.2 | 60.6 | 17.1 | 1.55 | 2.07 |
| B6  | 12 | 42.0 | 18.1 | 81.3 | 20.6 | 1.59 | 1.21 |
| B7  | 12 | 50.0 | 6.4  | 68.8 | 13.7 | 1.45 | 1.17 |
| B8  | 12 | 51.2 | 38.4 | 78.9 | 14.1 | 1.60 | 1.82 |
| B9  | 13 | 6.0  | 31.5 | 68.8 | 5.8  | 1.77 | 3.58 |
| B10 | 13 | 19.2 | 50.0 | 87.2 | 10.2 | 1.70 | 0.83 |
| B11 | 13 | 23.2 | 32.4 | 88.5 | 16.6 | 1.81 | 1.46 |
| B12 | 13 | 25.6 | 63.3 | 29.4 | 6.1  | 2.19 | 1.66 |
| B13 | 13 | 30.0 | 26.8 | 73.1 | 11.3 | 2.46 | 2.89 |
| B14 | 13 | 31.2 | 50.7 | 39.8 | 17.9 | 1.85 | 1.16 |
| B15 | 13 | 40.4 | 17.2 | 73.4 | 9.6  | 1.22 | 2.24 |
| B16 | 13 | 50.0 | 26.0 | 62.6 | 20.8 | 1.60 | 1.38 |
| B17 | 13 | 54.0 | 59.2 | 89.2 | 20.9 | 1.71 | 0.44 |
| B18 | 13 | 55.6 | 52.5 | 64.7 | 29.3 | 1.50 | 0.28 |
| B19 | 13 | 57.2 | 15.5 | 34.3 | 17.9 | 1.70 | 0.69 |
| B20 | 13 | 57.6 | 13.4 | 60.8 | 6.6  | 1.00 | 2.13 |
| B21 | 14 | 12.0 | 7.9  | 87.2 | 11.1 | 1.71 | 2.00 |
| B22 | 14 | 26.0 | 8.9  | 77.9 | 7.7  | 1.64 | 2.31 |
| B23 | 14 | 26.4 | 23.8 | 79.1 | 23.7 | 1.67 | 0.78 |
| B24 | 14 | 40.8 | 32.9 | 19.0 | 16.1 | 1.45 | 2.43 |
| B25 | 14 | 41.6 | 56.2 | 29.9 | 8.4  | 1.60 | 1.57 |
| B26 | 14 | 57.2 | 31.2 | 58.9 | 22.2 | 1.62 | 0.66 |
| B27 | 15 | 2.8  | 9.3  | 73.5 | 16.0 | 1.93 | 1.09 |
| B28 | 15 | 3.2  | 5.8  | 86.8 | 9.6  | 3.03 | 2.31 |
| B29 | 15 | 4.0  | 41.7 | 77.2 | 16.8 | 2.41 | 0.74 |
| B30 | 15 | 14.8 | 44.9 | 57.5 | 5.6  | 3.07 | 1.11 |
| B31 | 15 | 27.6 | 53.3 | 88.4 | 20.2 | 1.76 | 0.38 |
| B32 | 15 | 34.0 | 30.7 | 80.2 | 28.2 | 1.77 | 0.48 |
| B33 | 15 | 36.4 | 7.8  | 88.4 | 22.1 | 1.33 | 1.65 |
| B34 | 15 | 43.2 | 58.3 | 81.0 | 28.3 | 1.33 | 0.34 |
| B35 | 15 | 46.4 | 50.8 | 55.7 | 13.0 | 1.33 | 1.00 |
| B36 | 15 | 48.4 | 34.0 | 39.3 | 21.3 | 1.65 | 0.59 |
| B37 | 16 | 14.8 | 15.5 | 70.8 | 29.9 | 1.72 | 0.13 |
| B38 | 16 | 48.4 | 38.3 | 66.9 | 27.5 | 1.68 | 0.30 |
| B39 | 17 | 2.4  | 54.7 | 70.3 | 17.8 | 1.49 | 0.60 |
| B40 | 17 | 2.8  | 61.3 | 64.9 | 11.0 | 2.17 | 0.67 |
| B41 | 17 | 10.8 | 53.4 | 83.2 | 17.4 | 1.76 | 0.67 |

The last step in the construction of the void catalogue is the determination of void parameters. We take the center and diameter of voids found by the previous steps as the center and diameter of the whole void. The ellipticity of the void is determined by the maximum value of all segments drawn through the center of the "parent" void and the centers of all "children" voids. The ellipticity $e$ is calculated by dividing the length of this maximal axis by the diameter of the "parent" void,

$$e = \frac{max(r_0 + r + a)}{2r_0}, \quad (3)$$

**Table 2c.** Void Catalogue ($M \leq -18.8$)

| No. | $\alpha$ | | $\delta$ | $R$ | $D_v$ | $e$ | $\varrho$ |
|---|---|---|---|---|---|---|---|
| | h | m | | $h^{-1}$ Mpc | $h^{-1}$ Mpc | | |
| C1  | 12 | 24.4 | 10.8 | 22.5 | 3.2  | 1.89 | 5.28 |
| C2  | 12 | 43.6 | 36.6 | 58.9 | 15.6 | 1.00 | 1.92 |
| C3  | 12 | 50.0 | 14.9 | 51.0 | 20.5 | 1.52 | 0.81 |
| C4  | 13 | 19.2 | 47.5 | 36.9 | 15.0 | 1.68 | 1.53 |
| C5  | 13 | 32.0 | 33.7 | 53.3 | 5.6  | 1.41 | 1.73 |
| C6  | 13 | 32.8 | 18.9 | 25.2 | 9.9  | 1.84 | 2.67 |
| C7  | 14 | 0.4  | 20.8 | 37.7 | 15.6 | 1.67 | 0.91 |
| C8  | 14 | 4.0  | 16.0 | 56.9 | 4.3  | 1.78 | 2.37 |
| C9  | 14 | 7.2  | 42.1 | 46.8 | 11.9 | 1.74 | 1.57 |
| C10 | 14 | 11.2 | 55.9 | 59.7 | 20.0 | 1.14 | 0.30 |
| C11 | 14 | 22.8 | 9.2  | 49.7 | 9.2  | 2.11 | 1.37 |
| C12 | 14 | 42.0 | 45.3 | 48.2 | 9.7  | 2.02 | 1.21 |
| C13 | 14 | 44.0 | 32.3 | 59.7 | 19.0 | 1.58 | 1.07 |
| C14 | 14 | 55.6 | 12.0 | 45.2 | 12.9 | 1.51 | 0.73 |
| C15 | 15 | 9.6  | 16.9 | 57.1 | 14.3 | 1.52 | 1.07 |
| C16 | 15 | 21.6 | 32.5 | 16.6 | 14.4 | 1.44 | 2.95 |
| C17 | 15 | 30.0 | 15.5 | 30.1 | 12.4 | 1.74 | 0.62 |
| C18 | 15 | 30.4 | 70.1 | 39.2 | 15.6 | 1.45 | 0.80 |
| C19 | 15 | 38.4 | 16.5 | 47.3 | 6.8  | 1.47 | 0.97 |
| C20 | 15 | 55.2 | 11.9 | 54.7 | 7.0  | 2.49 | 0.80 |
| C21 | 16 | 9.6  | 44.0 | 56.3 | 10.7 | 1.96 | 1.17 |
| C22 | 16 | 15.6 | 49.2 | 50.2 | 15.6 | 1.58 | 0.83 |
| C23 | 16 | 21.2 | 39.0 | 39.3 | 22.8 | 1.50 | 0.43 |
| C24 | 16 | 29.2 | 28.7 | 59.6 | 19.0 | 1.70 | 0.33 |
| C25 | 17 | 21.6 | 60.0 | 57.5 | 7.5  | 1.00 | 0.66 |

$r_0$ is the radius of the "parent" void, $r$ is the radius of any other void, and $a$ is the distance of centers between this void and the "parent" void. The ellipticity parameter describes the deviation of the real void from an exact spherical shape.

The final void catalogue is presented in Table 2. Voids are listed separately for three absolute magnitude limits of galaxies used in our void definition. Voids found for the brightest galaxy limit are labeled A1 through A37, voids found for the next two magnitude limits B1 through B41 and C1 through C25. We give equatorial coordinates of void centers, distance $R$, void diameter $D_v$, ellipticity $e$, and the mean density $\varrho$ at the location of the void center. The latter will be explained in the next section. The limiting magnitude of galaxies used in the void definition is given in the header of Table 2. Some voids from our void catalogue (Table 2a) are plotted in Figure 4. The circles shown are intersections of the spherical voids with intermediate planes defined by z=10, 30, 50 and 70 $h^{-1}$ Mpc, respectively. Due to projection effects those "voids" are not completely empty.

**Table 3.** Overlapping Voids. Column (1): void number in the sample of brighter galaxies (cf. Table 2.). Column (2): void number in the sample of fainter galaxies. The overlapping degree (see Section 4.2) with the void from brighter sample is given in parenthesis.

| (1) | (2) |
|---|---|
| A2  | B17(25) |
| A3  | B4(22)  B8(45) |
| A4  | B4(38)  B6(49) |
| A5  | B11(45) |
| A6  | B19(68) |
| A9  | B17(93) |
| A12 | B13(65)  B16(80)  B23(50)  B26(44) |
| A16 | B22(21)  B27(30) |
| A19 | B17(30)  B18(50)  B29(86)  B31(72)  B34(62) |
| A24 | B33(54) |
| A25 | B26(26)  B27(41)  B32(46)  B33(29)  B37(100) |
|     | B38(23) |
| A30 | B34(45) |
| A37 | B41(74) |
| A6  | C4(20)  C6(67)  C7(68) |
| A12 | C13(55) |
| A19 | C10(35) |
| A25 | C15(53)  C20(85)  C24(61) |
| B2  | C1(52) |
| B5  | C3(36) |
| B14 | C4(84)  C9(36) |
| B16 | C13(28) |
| B18 | C10(90) |
| B19 | C6(42)  C7(76) |
| B24 | C6(31)  C16(81) |
| B26 | C13(92)  C15(24) |
| B35 | C21(39)  C22(55) |
| B36 | C23(77) |
| B37 | C24(29) |
| B38 | C21(44)  C24(51) |

### 4.4. The environment of voids

To determine the density of the environment of voids we calculated the density field for the cubic samples used in our void determination. As input data we used our absolute magnitude limited galaxy samples. The density field (number density of galaxies of all magnitudes in the respective magnitude interval) was calculated using Gaussian smoothing with a dispersion of 8 $h^{-1}$ Mpc for a $32^3$ grid. Experiments with numerical models (Frisch et al. 1994) show that this smoothing length characterizes the density field on supercluster scales. In Table 2 we give the density value at the location of the void center in units of the mean density $\bar{\varrho}$. These density values were used to divide the voids into three environmental density classes of high ($\varrho \geq 2\bar{\varrho}$), medium ($\bar{\varrho} \leq \varrho < 2\bar{\varrho}$) and low ($\varrho < \bar{\varrho}$) density.

**Table 4a.** Mean void diameters. Dependence on luminosity. Diameters are given with $\pm$ rms errors. $\sigma_v$ is the dispersion of void diameters in each sample. Mean diameters, typed with bold face at the end of each column of absolute magnitude limits correspond to complete samples for a given $L$.

| $L$ | Type | $M \leq -18.8$ | | $M \leq -19.7$ | | $M \leq -20.3$ | |
|---|---|---|---|---|---|---|---|
| | | $\langle D_v \rangle \pm \epsilon_v$ | $\sigma_v$ | $\langle D_v \rangle \pm \epsilon_v$ | $\sigma_v$ | $\langle D_v \rangle \pm \epsilon_v$ | $\sigma_v$ |
| $h^{-1}$ Mpc | | $h^{-1}$ Mpc | $h^{-1}$ Mpc | $h^{-1}$ Mpc | $h^{-1}$ Mpc | $h^{-1}$ Mpc | $h^{-1}$ Mpc |
| 30 | all | $8.9 \pm 1.3$ | 3.7 | $11.8 \pm 1.8$ | 4.0 | | |
| | E | $13.4 \pm 2.0$ | 4.0 | | | | |
| 45 | all | $12.3 \pm 1.0$ | 4.1 | $14.6 \pm 0.8$ | 3.1 | $17.1 \pm 0.9$ | 2.1 |
| | E | $15.7 \pm 1.4$ | 3.6 | $18.9 \pm 1.0$ | 1.8 | | |
| 60 | all | **$14.0 \pm 0.9$** | **4.0** | $16.4 \pm 0.8$ | 3.4 | $18.5 \pm 1.3$ | 3.0 |
| | E | **$17.7 \pm 1.3$** | **4.2** | $20.8 \pm 1.3$ | 2.8 | | |
| 75 | all | | | $17.1 \pm 1.1$ | 5.5 | $21.4 \pm 2.3$ | 7.4 |
| | E | | | $23.0 \pm 1.8$ | 5.7 | $31.1 \pm 5.7$ | 9.5 |
| 90 | all | | | **$18.8 \pm 0.9$** | **5.3** | $23.8 \pm 1.6$ | 6.8 |
| | E | | | **$27.3 \pm 2.3$** | **7.8** | $37.7 \pm 3.7$ | 8.4 |
| 105 | all | | | | | $24.8 \pm 1.4$ | 6.9 |
| | E | | | | | $34.1 \pm 3.5$ | 9.0 |
| 120 | all | | | | | **$25.7 \pm 1.3$** | **6.4** |
| | E | | | | | **$36.4 \pm 3.4$** | **8.4** |

Examples of equidensity contours for superclusters (high density regions with $\varrho \geq 2\bar{\varrho}$) and the Northern Local Supervoid (low density region with $\varrho < \bar{\varrho}$) are shown in Figure 1b.

### 4.5. The analysis of the void catalogue

In this section we compare voids defined by galaxies of different magnitude limits. About one third of the voids in sample A can be identified also in sample B. To make such comparison easier we give in Table 3 a list of overlapping voids. Because barely overlapping voids are of no interest, void pairs with overlapping degrees less than 20 per cent are not included in the Table. In the first part of the Table we list the names of voids of sample B overlapping with voids of sample A, in the next two parts we give similar information for voids of sample C. In parenthesis we give the overlapping degree as defined above. Only one pair of voids in Table 3 overlaps 100 per cent, indicating that void B37 lies completely inside of void A25.

Tables 2 and 3 show that in sample B the mean diameter of voids is smaller than in sample A, and in sample C it is smaller than in sample B. In the example mentioned above the smaller void B37 with $D_v = 29.9\ h^{-1}$ Mpc (see Table 2b) lies completely inside the larger void A25 ($D_v = 39.5\ h^{-1}$ Mpc) of the brighter sample. Other examples are B17 ($D_v = 20.9\ h^{-1}$ Mpc) which is almost completely inside void A9 ($D_v = 21.3\ h^{-1}$ Mpc and $\theta = 93\ \%$), and C10 ($D_v = 20.9\ h^{-1}$ Mpc) which is almost inside B18 ($D_v = 29.3\ h^{-1}$ Mpc and $\theta = 90\ \%$). In these cases systems of faint galaxies enter into the void, and make its diameter smaller.

In other cases a large void in sample A has been split into several smaller voids in samples B or C. For example A6 ($D_v = 20.6\ h^{-1}$ Mpc) is split into C6 ($D_v = 9.9\ h^{-1}$ Mpc and $\theta = 67\ \%$) and C7 ($D_v = 15.6\ h^{-1}$ Mpc and $\theta = 68\ \%$). Those two voids defined by the faintest galaxies also show strong overlapping with void B19 ($D_v = 17.9\ h^{-1}$ Mpc). B19 itself lies partly inside void A6 ($\theta = 68\ \%$). Many further examples of such an overlapping hierarchy can be found from Table 3. In these cases one or more faint galaxy systems split the void into several sub-voids. Here we see a full analogy with the structure of supervoids which are split by systems of poor clusters and galaxies into smaller interconnecting sub-voids.

## 5. Diameters of Voids

### 5.1. Determination of void diameters

For a more detailed investigation of the dependence of the void properties, specifically of their diameter, on the type of objects used to define the void, the same void analysis as described in Section 4.1 was applied to samples of clusters of galaxies, galaxies of all morphological types and separately to eliptical galaxies. We used cube sizes $L = 120$, 90, 60 and 30 $h^{-1}$ Mpc and the corresponding absolute magnitude limits $M = -20.3$, $-19.7$, $-18.8$ and $-17.3$. To investigate the sensitivity of the results to the resolution parameter $k$ we used $k = 16$, 24, 32, 40, 48, 56 and 64 with resulting cell sizes in the range of 1.88 to 0.47 $h^{-1}$ Mpc for $L = 30$ $h^{-1}$ Mpc, 3.75 to 0.94 $h^{-1}$ Mpc for $L = 60$ $h^{-1}$ Mpc, and 7.5 to 1.88 $h^{-1}$ Mpc for $L = 120$ $h^{-1}$ Mpc. Additionally we used subsamples with $L = 45$, 75 and 105 $h^{-1}$ Mpc, and respective absolute magnitude limits.

For each galaxy sample we determined the void sample using the same void selection procedure as discussed above. We kept only voids lying inside a sphere with the center at the origin of the coordinate system and radius equal to the cube size $L$. Voids located closer to the sample boundaries than the radius of the void were eliminated. Mean void diameters were calculated in two steps. First the mean value was calculated from voids found with the empty sphere method using the corresponding resolution parameter $k$. In the second step the mean value for all $k$ was calculated.

To study the sensitivity of our results to details of the elimination procedure we calculated also mean void diameters using a constant distance from the sample boundary. In first approximation, we took the mean radius of the void to be 10 $h^{-1}$ Mpc, the mean void radius of galaxy samples. We found that mean void diameters are almost the same for variable and constant distant from sample boundaries used to reduce void catalogues. We carried out analogous studies for a cluster sample with cube size $L = 120$ $h^{-1}$ Mpc and three different distances $\Delta = 10$, 15 and 20 $h^{-1}$ Mpc from the cube borders to eliminate voids. The results do not show any significant differences in the number and mean diameter of voids depending on the distance to the cube borders.

To reduce multiple counting of voids we used the following procedure: look for all pairs of empty spheres, calculate the overlapping degree $\theta$ and, if $\theta > 50$ %, then the sphere with the smaller radius is discarded. For comparison, we also calculated the mean diameter of all voids, including all overlapping voids. The results show that the distribution of void diameters in both samples are rather similar, and thus the elimination of overlapping voids only slightly reduces the mean void diameter. The deviation increases with increasing resolution parameter $k$ and reaches up to 10 %. This deviation is much smaller than the spread of mean diameters inside a particular void sample.

The overlapping void elimination procedure used in this section is different from the respective procedure used to construct our void catalogues in Section 4.3. For this reason mean void diameters for full samples given in Table 4a do not exactly coincide with respective data for all galaxies given in Table 4b. These differences, although small, demonstrate that in comparing void sizes it is necessary to use data obtained with identical void definition algorithms.

### 5.2. Results

Mean void diameters of galaxy samples for all cube sizes used are given in Table 1, separately for voids defined by all galaxies and by elliptical galaxies, as well as by poor clusters of galaxies. Nearby galaxy samples contain both faint and bright galaxies, thus for nearby samples it is possible to calculate void diameters using several absolute magnitude limits. In Table 4a the mean diameters of voids, $\langle D_v \rangle$, are given for various absolute magnitude limits, $M_0$, and for different cube sizes, $L$. We computed the scatter of mean void diameters, $\sigma_v$. From this we estimated errors of the mean void diameters, $\epsilon_v$, as

$$\epsilon_v = \sigma_v / \sqrt{N_v}, \qquad (4)$$

where $N_v$ is the number of voids used in the determination of the mean value. Void diameter scatter and mean diameter errors are also given in Tables 1 and 4.

In Figure 6 we show the cumulative distribution of void diameters for some samples. In Figure 6a the cumulative distribution of diameters of voids defined by poor clusters within the 120 $h^{-1}$ Mpc cubic sample is plotted for different resolutions $k$. We see that the scatter of individual curves is rather small and there is no systematic dependence of the distribution on $k$. Similar plots for galaxy samples show no systematic differences for $k$. Thus we can use the mean value for all $k$, as a reliable distribution curve of void diameters.

The cumulative frequency distribution of void diameters in all galaxy and elliptical galaxy samples are shown in Figures 6b and 6c, respectively. The void diameter distribution is calculated for galaxy samples with absolute magnitude limits $-18.8$, $-19.7$ and $-20.3$.

### 5.3. Dependence on morphology and luminosity

Mean void diameters of samples of various cube sizes $L$ for galaxies of different morphological type, and different luminosity are listed in Tables 1 and 4. In some samples the number of elliptical galaxies was too small to calculate the mean void diameter; in these cases no void diameter value is given. In the case of the cube sizes $L = 60$ $h^{-1}$ Mpc, 90 $h^{-1}$ Mpc and 120 $h^{-1}$ Mpc the values with the absolute magnitude limit corresponding to these distances are printed bold face in Table 4a. A graphical presentation of the results is given in Figure 7.

The comparison of diameters of voids defined by galaxies of different morphological type and magnitude shows that voids defined by luminous galaxies and galaxies of early morphological type tend to be larger than voids defined by fainter galaxies and galaxies of all types. The reason for this can be seen in Figure 5, which shows that intrinsically fainter galaxies form faint systems in voids delineated by bright galaxies. Also, faint galaxies populate peripheral regions of galaxy systems. Therefore faint galaxies form structures with smaller voids. Similarly we can see in Figure 4 that elliptical galaxies are concentrated to the centers of galaxy clusters or groups and to the ridges of systems of galaxies, and are absent in low-density peripheral regions of systems of galaxies. This effect is known from earlier studies as the large-scale morphological segregation of galaxies (Mo et al. 1992).

**Fig. 6.** Panel a) shows the distribution of diameters of voids defined by clusters of galaxies for different resolution parameter $k$. Panels b) and c) show the distribution of diameters (mean for different resolution parameter $k$) of voids defined by all galaxies and elliptical galaxies; different lines give the distribution of void diameters defined by galaxies of absolute magnitude, brighter than $M = -18.8$, $-19.7$ and $-20.3$. In panels d) – f) we give the distribution of diameters of voids located in different environment (for classification criteria we refer to Section 5.5), separately for galaxies of different absolute magnitude limit.

A comparison of void distributions determined by galaxies (Figures 6b and 6c) and by clusters of galaxies (Figure 6a) suggests a dependence of mean void diameters on the absolute magnitude and morphology of galaxies. The distribution curves for elliptical galaxies are shifted to the right compared with the curves for galaxies of all types indicating that diameters of voids outlined by elliptical galaxies are larger. Comparing Figure 6a and 6c we recognize that the distribution of void diameters for the bright elliptical galaxies resembles strongly that for Zwicky clusters.

Figure 6 shows that the scatter of void diameters also depends on the type of object used in the void determination. The cumulative void diameter distribution for galaxies is steeper than for clusters, indicating that the range of void diameters is smaller than for clusters. This result is confirmed by the corresponding values ($\sigma_v$ for L=60, 90 and 120 $h^{-1}$ Mpc) in Table 1. The relative rms scatter (calculated as $\sigma_v$ over $\langle D_v \rangle$ from Table 1) of void diameters for clusters amounts to 32 % whereas the mean relative scatter of galaxy defined voids for all galaxy samples in Table 1 is 27 %.

The overall range of mean void diameters defined by all galaxies ranges from 9 to 25 $h^{-1}$ Mpc, and from 13 to 36 $h^{-1}$ Mpc for voids defined by elliptical galaxies. This range is much larger than the statistical error of mean

**Fig. 7.** Panel a) shows the dependence of mean diameters of voids defined by all galaxies with absolute magnitude limits $M = -18.8, -19.7$ and $-20.3$ on the cube size of the sample. Panel b) shows the dependence of mean diameters of voids defined by elliptical galaxies with absolute magnitude limits $M = -18.8, -19.7$ and $-20.3$ on the cube size of the sample. In the case of filled symbols the luminosity limit corresponds to a complete sample for a given size. Error bars indicate the rms error of void diameters.

void diameters, thus the effect is statistically significant. When we compare voids defined by galaxies of different absolute magnitude inside the same parent sample (i.e. values in the same row of Table 4a), we get a smaller range. But again the range of void diameters is much larger than the statistical errors. In section 5.6 we will argue that the differences in mean void diameters are not only due to differences in number density of galaxies in samples.

### 5.4. Dependence on the sample size

Next, we consider the dependence of the mean void size on the sample size. Each column in Table 4a gives mean void diameters for some subsamples of a given absolute magnitude limited sample. The results listed in Table 4a are plotted in Figure 7, for all galaxies in Figure 7a, and for elliptical galaxies in Figure 7b. Vertical bars indicate the mean statistical error of void diameters.

Figure 7 demonstrates that there is a slow but continuous increase of the mean void size with the sample size. In Figure 7 void diameters of samples taken at absolute magnitude corresponding to the completeness limit for a given sample size are plotted with solid symbols, in Table 4a respective void diameters are printed bold-face. We see, that if we change the two parameters (sample size and absolute magnitude limit) simultaneously, then we see a more rapid increase of void sizes with sample size. This effect, the *self-similarity* of voids, has been described by several authors (EEG, KF, HB).

Due to the dependence of the void size on the luminosity of galaxies an increase of the void size with increasing distance is expected since samples taken at larger distance use a higher luminosity limit. Our analysis shows that the luminosity effect only partly explains the observed self-similarity of voids. Another possibility is that the effect is due to biases because in samples taken in cubes of larger size voids of larger diameter can be detected. In this case the void diameter distribution must have a tail towards large diameters. Figures 6b and 6c show that this is not the case. We are left with the possibility that void sizes depend on the large-scale environment: nearby samples are located in the Local supercluster, and with increasing distance our samples enter more into the supervoid region which is a region of larger voids as explained in the next section.

### 5.5. Dependence on the environment

To investigate the dependence of void sizes on the environment we have calculated the smoothed density field

(Gaussian smoothing with dispersion 8 $h^{-1}$ Mpc) for the three cubic samples used to generate the void catalogues in Table 2a, 2b and 2c (cf. Section 4.4). To calculate the cumulative distribution of void diameters we have divided the voids in our void catalogues into three classes – voids located in high, medium and low density environment. The limiting density values between high- and medium density, and medium and low density environment are 2 and 1 in units of the mean density $\bar{\varrho}$, respectively. In Figure 8 the void diameters are plotted as a function of the smoothed density at the location of the void center. Mean diameters of voids $\langle D_v \rangle$ for these three environment classes are given in Table 4b, we also give the scatter $\sigma_v$ and statistical error of the diameters. The cumulative distribution of void diameters for these environmental classes is plotted in the panels d, e and f of Figure 6.

Our calculations show that voids located in a high-density environment are smaller than voids located in a low-density environment. The mean diameters differ by a factor of $\approx 2$. Fainter galaxies determine smaller voids in all environments; the smallest voids defined by faint galaxy systems in high-density environments have a mean diameter of only 8 $h^{-1}$ Mpc. The effect is much larger than the statistical errors of void diameters or from the change in number densities (see next section).

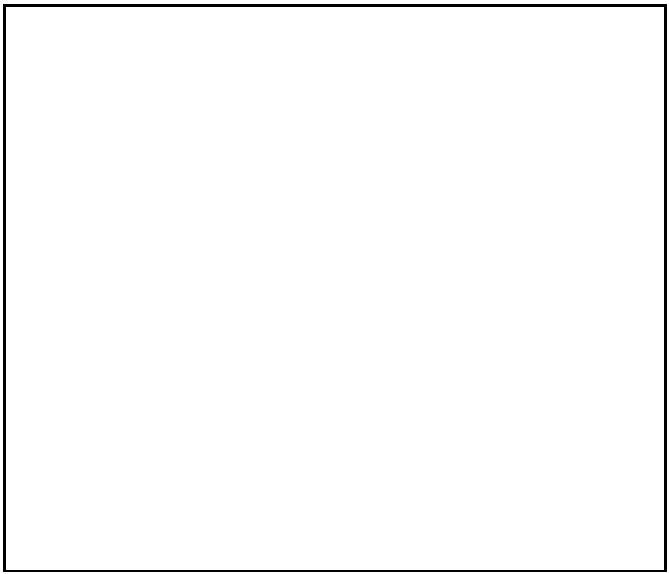

**Fig. 8.** The dependence of void diameters, $D_v$, on the density at the void center, $\varrho$. Voids from our catalogue in Table 2a (A1 – A37) are marked with open squares, voids from catalogue Table 2b (B1 – B41) with open triangles, and voids from catalogue Table 2c (C1 - C25) with open circles.

The conclusion on the dependence of mean void diameters on the large-scale environment is an important result of this study. It means that in high-density regions not only systems of galaxies are richer (this follows from the definition of high-density regions) but also systems of galaxies are spaced more closely.

**Table 4b.** Mean void diameters. Dependence on environment. $\langle D_v \rangle$ and $\sigma_v$ are given in units $h^{-1}$ Mpc.

| Env | $M \leq -18.8$ | | $M \leq -19.7$ | | $M \leq -20.3$ | |
|---|---|---|---|---|---|---|
| | $\langle D_v \rangle \pm \epsilon_v$ | $\sigma_v$ | $\langle D_v \rangle \pm \epsilon_v$ | $\sigma_v$ | $\langle D_v \rangle \pm \epsilon_v$ | $\sigma_v$ |
| h | $7.9 \pm 2.3$ | 4.5 | $9.2 \pm 1.2$ | 4.0 | $16.4 \pm 1.2$ | 3.9 |
| m | $12.3 \pm 1.3$ | 3.8 | $14.4 \pm 1.4$ | 5.2 | $19.3 \pm 1.4$ | 5.0 |
| l | $14.6 \pm 1.5$ | 5.2 | $21.4 \pm 1.5$ | 6.0 | $30.8 \pm 1.4$ | 5.2 |
| all | $12.7 \pm 1.0$ | 5.2 | $15.6 \pm 1.1$ | 7.3 | $22.5 \pm 1.3$ | 7.8 |

### 5.6. Comparison with Poisson data

A dependence of void diameters on the absolute luminosity of galaxies is expected since the number density of bright galaxies is lower than that of fainter ones. Therefore, in order to investigate the significance of the results of the previous sections, we have generated a number of random samples of different size and number of particles which cover the range of sizes and number densities of real samples. We computed the void size distribution, and its scatter, using the same algorithm as for real samples. Our experiments show that the mean diameter of voids in a Poisson sample is

$$D_{Poiss} \approx 3 R_0, \quad (5)$$

where the Poisson radius, $R_0$, for a cubic sample of size $L$ is defined as

$$R_0 = \left( \frac{3}{4 \pi N} \right)^{1/3} L, \quad (6)$$

where $N$ is the number of objects in the cubic sample.

The void diameters for a number of samples are plotted in Figure 9 as a function of the number of particles in cubes of size 30, 60 and 90 $h^{-1}$ Mpc. As a comparison, we show the equivalent void diameters as a function of the number of particles for Poissonian samples. In the $\log N - \log D_v$-diagram of Figure 9 the dependence for Poisson samples is given by straight lines of slope $-1/3$. It can be seen from Figure 9 that for full samples (samples with faintest galaxies included and therefore maximum $N$) the mean void diameter exceeds the mean diameter of voids in Poisson samples by a factor of $\approx 1.5$. Galaxies of increasingly higher luminosity (i.e. smaller $N$) define voids of larger diameter. Elliptical galaxy samples and samples of brightest galaxies (i.e. minimum $N$) define voids with diameters, approximately equal to diameters of voids in Poisson samples. The dependence of void diameters on the number of particles has, however, a different slope, considerably lower than for Poisson samples.

**Fig. 9.** Void diameters plotted as a function of the number of particles, $N$, for samples taken in cubes of size 30, 60, and 90 $h^{-1}$ Mpc. Straight lines are for Poisson samples of respective size. Open symbols are for samples of galaxies of all types, filled symbols for samples of elliptical galaxies.

The second difference of real voids from Poissonian ones lies in the distribution of void diameters – the mean spread of void diameters in Poisson samples is 0.19 (in units of the mean void diameters) whereas in real samples the spread is much larger, 0.27 for voids defined by galaxies and 0.32 for cluster-defined voids (cf. 27% and 32% determined in section 5.3). The spread is largest in bright galaxy and cluster samples (cf. Table 1) where mean diameters of voids are close to the diameters of voids in Poisson samples with the same number density). We therefore conclude that statistical properties of the identified voids are different from "voids" found in Poissonian samples. We note that the relative spread of void diameters is one of the statistics used by Dekel et al. (1992) in the comparison of real data with simulations and toy models.

Similar differences between real void diameters and diameters of voids in Poisson samples were found earlier by EEGS, for characteristic void diameters defined by void probability functions. For our study it is important that these differences increase when we consider samples of fainter galaxies. We conclude that the void diameter statistics as well as the void probability function characterize certain properties of the galaxy distribution, this analysis is especially sensitive to properties of the distribution of galaxies in low-density regions.

### 5.7. The influence of the incompleteness of observational data

In order to investigate the impact of the incompleteness of data on our void diameter statistics we proceed as follows. In fields where data are complete up to 15.5 we have randomly removed part of the galaxies in the apparent magnitude interval $14.5 - 15.5$ to get samples which are complete relative to full samples 30 %, 50 %, and 70 %. Using these diluted samples we have repeated the void analysis. Results of the void analysis show that for the absolute magnitude interval $M \leq -18.8$ for all cube sizes $L = 30, 45, 60\ h^{-1}$ Mpc there is no systematic change of mean void diameters for any of the dilution degree applied. For brighter absolute magnitude limits $M \leq -19.7$ and $-20.3$ there exists a marginal increase of mean void diameters for larger samples ($L > 60\ h^{-1}$ Mpc): by 12 %, 5 % and 2 % for dilution levels 30 %, 50 % and 70 % of the full sample, respectively. In samples of smaller size there is practically no change of mean diameters.

These calculations show that the incompleteness influences our void diameters insignificantly. Since we are interested basically in comparing of void diameters for samples of different absolute magnitude, we conclude that the differential effect is negligible.

## 6. Discussion

In the classical treatment of voids the dependence of void sizes on the morphology and luminosity of galaxies is ignored. This approach is equivalent to the assumption that the void distribution is non-hierarchical, as in the HDM scenarios of structure formation. The analysis of voids in numerical models by Frisch et al. (1994) has shown that in this case voids defined by simulated galaxies have mean sizes comparable to sizes of cluster-defined voids in most simulated and observed cluster samples. This sort of void distribution has been tacitly adopted by most previous void searches.

In this classical approach voids define a certain scale in the Universe that is almost independent on the type of objects used in void definition. Our study has shown that in the observed samples there is no preferred scale of galaxy-defined voids – all scales between $\approx 5\ h^{-1}$ Mpc and $\approx 50\ h^{-1}$ Mpc are present in galaxy-defined voids, depending on the chosen absolute magnitude limit and the large-scale environment, see Figure 9. This figure shows that the mean void diameters in observed samples increase continuously with decreasing number of particles.

Similarly, clusters of galaxies of different richness define voids of different scale, from $\approx 40\ h^{-1}$ Mpc for Zwicky clusters of galaxies to $\approx 100\ h^{-1}$ Mpc for Abell clusters of all richness classes.

The concentration of bright galaxies, especially bright elliptical galaxies, to central parts of galaxy systems is well-known from earlier studies. In particular, the study of the correlation function and the power spectrum of galaxies has shown that differences in the spatial distribution of galaxies exist for intrinsically bright galaxies, $M \leq -20$ (Hamilton 1988, Einasto 1991, Gramann and

Einasto 1992). Our study of void diameters has demonstrated that differences exist in the distribution of fainter galaxies, too. Vogeley et al. (1994b) came to a similar conclusion.

The principal mechanism which creates the hierarchy in void distribution is the presence of multi-branching systems in the galaxy distribution. Bright galaxies and poor clusters form systems which split a large cluster-defined void into smaller units. These systems have smaller branches made of fainter galaxies which define voids of even smaller scale. The fainter the galaxies we considered, the finer are systems they form. Thus galaxies of different magnitude and morphological type define voids of different sizes. Smallest sizes have voids defined by dwarf galaxies in high-density regions of superclusters, their mean diameters are only 8 $h^{-1}$ Mpc (Table 2).

The galaxy sample used in this paper is not complete to galaxies fainter than $M = -18.8$. From the above discussion, one could extrapolate that the hierarchy of voids continues to fainter galaxies and that fainter galaxies define voids of still smaller sizes. Available data on the distribution of dwarf galaxies do not support this hypothesis (Bothun et al. 1993, Szomoru et al. 1994, Hopp 1994, and references therein about the search of dwarf galaxies in voids). Very faint galaxies do not form a smooth population in voids delineated by brighter galaxies, but are located in the outskirts of systems of brighter galaxies or form systems consisting only of faint galaxies. These systems determine the lower limit of the hierarchical structure.

On the other hand, structures formed by superclusters of galaxies have a certain characteristic scale $\approx 130$ $h^{-1}$ Mpc (EETDA) which corresponds to the upper limit of the hierarchy of structures.

The study of the Bootes void by Szomoru et al. (1994) has also shown a filamentary structure in this void. We see that both supervoids are similar in their internal structure.

The idea of the presence of a hierarchy of systems of galaxies has been advocated long ago by de Vaucouleurs (1970). The confirmation of the hierarchical picture of the distribution of systems of galaxies and voids is the main result of this study.

## 7. Conclusions

We have studied the structure of the Northern Local void, a cluster-defined supervoid between the Local, Coma, and Hercules Superclusters. The main results of our empirical study are:
1) Starting from the observation that the Northern Local void is not empty we established the presence of a hierarchical fine structure in the form of a network of systems of galaxies and poor clusters in this supervoid.
2) We presented a catalogue of voids defined by galaxies of different luminosity.
3) We demonstrated that the whole void-filament structure is hierarchical: voids defined by poor clusters of galaxies and by bright galaxies contain smaller voids defined by fainter galaxies. Void sizes depend crucially on the properties of galaxies considered. Therefore the type of objects defining voids must be specified if void properties are investigated.
4) Voids located in high-density environment in superclusters are smaller than voids located in low-density environment.

These are observed properties of the distribution of galaxies, and eventually, a realistic galaxy formation scenario has to explain them.

*Acknowledgements.* We thank Drs Martha Haynes and Bianca Garilli for providing computer files of redshift measurements, Marcio Maia for providing unpublished redshift data, and Peter Shaver for providing unpublished redshift data on radio galaxies. JE and ME thank the staff of European Southern Observatory, Potsdam Institute for Astrophysics, and Göttingen University Observatory for support and very stimulating atmosphere that made this collaboration possible. Fruitful discussions with Neta Bahcall, Mirt Gramann, and Enn Saar are acknowledged. This study was supported by Estonian Science Foundation grant 182, International Science Foundation grant LDC 000, and a grant of the Göttingen Academy of Sciences. We thank the anonymous referee for constructive criticism.